\documentclass[prl,superscriptaddress,showpacs,amssymb,amsmath,amsfonts,aps,twocolumn,secnumarabic,floatfix]{revtex4}
\usepackage{graphics}
\usepackage{epsfig}
\begin{document} 
\bibliographystyle{try}
\topmargin -.9cm 

 \title{Experimental study of exclusive $^2$H$(e,e^\prime p)n$ reaction mechanisms at high $Q^2$}  

\date{\today}

\newcommand*{\ANL}{Argonne National Laboratory, Argonne, IL 60439}
\affiliation{\ANL}
\newcommand*{\ASU}{Arizona State University, Tempe, Arizona 85287-1504}
\affiliation{\ASU}
\newcommand*{\UCLA}{University of California at Los Angeles, Los Angeles, California  90095-1547}
\affiliation{\UCLA}
\newcommand*{\CSU}{California State University, Dominguez Hills, Carson, CA 90747}
\affiliation{\CSU}
\newcommand*{\CMU}{Carnegie Mellon University, Pittsburgh, Pennsylvania 15213}
\affiliation{\CMU}
\newcommand*{\CUA}{Catholic University of America, Washington, D.C. 20064}
\affiliation{\CUA}
\newcommand*{\SACLAY}{CEA-Saclay, Service de Physique Nucl\'eaire, 91191 Gif-sur-Yvette, France}
\affiliation{\SACLAY}
\newcommand*{\CNU}{Christopher Newport University, Newport News, Virginia 23606}
\affiliation{\CNU}
\newcommand*{\UCONN}{University of Connecticut, Storrs, Connecticut 06269}
\affiliation{\UCONN}
\newcommand*{\ECOSSEE}{Edinburgh University, Edinburgh EH9 3JZ, United Kingdom}
\affiliation{\ECOSSEE}
\newcommand*{\EMMY}{Emmy-Noether Foundation, Germany}
\affiliation{\EMMY}
\newcommand*{\FU}{Fairfield University, Fairfield CT 06824}
\affiliation{\FU}
\newcommand*{\FIU}{Florida International University, Miami, Florida 33199}
\affiliation{\FIU}
\newcommand*{\FSU}{Florida State University, Tallahassee, Florida 32306}
\affiliation{\FSU}
\newcommand*{\GWU}{The George Washington University, Washington, DC 20052}
\affiliation{\GWU}
\newcommand*{\ECOSSEG}{University of Glasgow, Glasgow G12 8QQ, United Kingdom}
\affiliation{\ECOSSEG}
\newcommand*{\ISU}{Idaho State University, Pocatello, Idaho 83209}
\affiliation{\ISU}
\newcommand*{\INFNFR}{INFN, Laboratori Nazionali di Frascati, 00044 Frascati, Italy}
\affiliation{\INFNFR}
\newcommand*{\INFNGE}{INFN, Sezione di Genova, 16146 Genova, Italy}
\affiliation{\INFNGE}
\newcommand*{\ORSAY}{Institut de Physique Nucleaire ORSAY, Orsay, France}
\affiliation{\ORSAY}
\newcommand*{\BONN}{Institute f\"{u}r Strahlen und Kernphysik, Universit\"{a}t Bonn, Germany}
\affiliation{\BONN}
\newcommand*{\ITEP}{Institute of Theoretical and Experimental Physics, Moscow, 117259, Russia}
\affiliation{\ITEP}
\newcommand*{\JMU}{James Madison University, Harrisonburg, Virginia 22807}
\affiliation{\JMU}
\newcommand*{\KYUNGPOOK}{Kyungpook National University, Daegu 702-701, South Korea}
\affiliation{\KYUNGPOOK}
\newcommand*{\MIT}{Massachusetts Institute of Technology, Cambridge, Massachusetts  02139-4307}
\affiliation{\MIT}
\newcommand*{\UMASS}{University of Massachusetts, Amherst, Massachusetts  01003}
\affiliation{\UMASS}
\newcommand*{\MOSCOW}{Moscow State University, General Nuclear Physics Institute, 119899 Moscow, Russia}
\affiliation{\MOSCOW}
\newcommand*{\UNH}{University of New Hampshire, Durham, New Hampshire 03824-3568}
\affiliation{\UNH}
\newcommand*{\NSU}{Norfolk State University, Norfolk, Virginia 23504}
\affiliation{\NSU}
\newcommand*{\OHIOU}{Ohio University, Athens, Ohio  45701}
\affiliation{\OHIOU}
\newcommand*{\ODU}{Old Dominion University, Norfolk, Virginia 23529}
\affiliation{\ODU}
\newcommand*{\PITT}{University of Pittsburgh, Pittsburgh, Pennsylvania 15260}
\affiliation{\PITT}
\newcommand*{\RPI}{Rensselaer Polytechnic Institute, Troy, New York 12180-3590}
\affiliation{\RPI}
\newcommand*{\RICE}{Rice University, Houston, Texas 77005-1892}
\affiliation{\RICE}
\newcommand*{\URICH}{University of Richmond, Richmond, Virginia 23173}
\affiliation{\URICH}
\newcommand*{\SCAROLINA}{University of South Carolina, Columbia, South Carolina 29208}
\affiliation{\SCAROLINA}
\newcommand*{\JLAB}{Thomas Jefferson National Accelerator Facility, Newport News, Virginia 23606}
\affiliation{\JLAB}
\newcommand*{\UNIONC}{Union College, Schenectady, NY 12308}
\affiliation{\UNIONC}
\newcommand*{\VT}{Virginia Polytechnic Institute and State University, Blacksburg, Virginia   24061-0435}
\affiliation{\VT}
\newcommand*{\VIRGINIA}{University of Virginia, Charlottesville, Virginia 22901}
\affiliation{\VIRGINIA}
\newcommand*{\WM}{College of William and Mary, Williamsburg, Virginia 23187-8795}
\affiliation{\WM}
\newcommand*{\YEREVAN}{Yerevan Physics Institute, 375036 Yerevan, Armenia}
\affiliation{\YEREVAN}
\newcommand*{\NOWOHIOU}{Ohio University, Athens, Ohio  45701}
\newcommand*{\NOWUNH}{University of New Hampshire, Durham, New Hampshire 03824-3568}
\newcommand*{\NOWUMASS}{University of Massachusetts, Amherst, Massachusetts  01003}
\newcommand*{\NOWMIT}{Massachusetts Institute of Technology, Cambridge, Massachusetts  02139-4307}
\newcommand*{\NOWURICH}{University of Richmond, Richmond, Virginia 23173}
\newcommand*{\NOWCUA}{Catholic University of America, Washington, D.C. 20064}
\newcommand*{\NOWECOSSEE}{Edinburgh University, Edinburgh EH9 3JZ, United Kingdom}
\newcommand*{\NOWGEISSEN}{Physikalisches Institut der Universitaet Giessen, 35392 Giessen, Germany}

\newcommand*{\NOWREGINA}{University of Regina, Regina, SK S4S0A2, Canada}

\author {K.S.~Egiyan} 
\thanks{Deceased}
\affiliation{\YEREVAN}
\affiliation{\JLAB}
\author {G.~Asryan} 
\affiliation{\YEREVAN}
\author {N.~Gevorgyan} 
\affiliation{\YEREVAN}
\author {K.A.~Griffioen} 
\affiliation{\WM}
\author {J.M.~Laget} 
     \email{laget@jlab.org}
     \thanks{Corresponding author.}
\affiliation{\JLAB}
\author {S.E.~Kuhn} 
\affiliation{\ODU}
\author {G.~Adams} 
\affiliation{\RPI}
\author {M.J.~Amaryan} 
\affiliation{\ODU}
\author {P.~Ambrozewicz} 
\affiliation{\FIU}
\author {M.~Anghinolfi} 
\affiliation{\INFNGE}
% ...
\collaboration{The CLAS Collaboration}
     \noaffiliation

\begin{abstract} 

The reaction $^2$H$(e,e^\prime p)n$ has been studied with full kinematic coverage 
for photon virtuality 
$1.75<Q^2<5.5$~GeV$^2$.  
Comparisons of experimental data with theory
 indicate that for very low values of neutron recoil momentum ($p_n<100$ MeV/c) 
the neutron is primarily a spectator and the reaction can be described by
the plane-wave impulse approximation.
For $100<p_n<750$ MeV/c  proton-neutron rescattering dominates the cross section, 
while $\Delta$ production followed by the $N\Delta \rightarrow NN$ transition is the 
primary contribution at higher momenta.

\end{abstract} 


\maketitle 

For high  virtuality of the exchanged photon, the $^2$H$(e,e^\prime p)n$ reaction is one of the simplest and best 
ways to investigate high-momentum components of the deuterium wave function, possible modifications to the internal 
structure of bound nucleons, and
the nature of short-range nucleon correlations. To date this reaction was studied only for
low $Q^2$ ($< 1$ GeV$^2$) at Saclay, NIKHEF, Mainz and Bates. A survey prior to 1990 can be found in 
Ref.~\cite{La91}. In general, the interpretation of these results suffered from large corrections due to 
final-state interactions (FSIs), meson exchange currents (MECs) and the intermediate $\Delta$ contribution. 

The  Continuous Electron Beam Accelerator Facility (CEBAF) at Jefferson Laboratory (JLab) has opened a new frontier in the study of 
$^2$H$(e,e^\prime p)n$ and other $(e,e^\prime p)$ reactions for $Q^2$ up to 6 GeV$^2$.  The first study of the exclusive 
$^2$H$(e,e^\prime p)n$ reaction at JLab has been carried out in  Hall A~\cite {paul}. The cross section was measured 
as a function of recoil momentum, $p_n$, up to 550 MeV/c in perpendicular kinematics
with $Q^2=0.68$~GeV$^2$.  At low ($<300$ MeV/c) recoil momentum, these data can be described to within 1-2$\sigma$ by the Plane Wave Impulse Approximation (PWIA),
while at higher momenta FSIs and the $\Delta$ contribution must be included. 
Two new experiments have been carried out at higher $Q^2$: the first in Hall A~\cite{Bo03} for $Q^2< 3.5$ GeV$^2$  and the 
second in Hall B~\cite{kim} for $1.75< Q^2< 5.5$ GeV$^2$ which is reported in this letter.
We have done a comprehensive study 
of the $^2$H$(e,e^\prime p)n$ exclusive reaction with full kinematic coverage, which allows us to identify the dominant mechanisms.

The experiment has been performed using the CEBAF Large-Acceptance Spectrometer (CLAS)~\cite{CLAS}, which consists of six sectors, each
functioning as an independent magnetic spectrometer. Each sector is instrumented with multi-wire drift chambers,
time-of-flight scintillator counters covering polar angles $8^{\circ}<\theta<143^{\circ}$, 
gas-filled threshold Cherenkov counters (CCs) and
lead-scintillator sandwich-type electromagnetic calorimeters (ECs) covering $8^{\circ}<\theta<45^{\circ}$. The CLAS was triggered on scattered electrons  
identified by a coincidence between EC and CC signals in a given sector.

A 5.761 GeV electron beam impinged on a 
target cell of liquid deuterium about 5 cm long and 0.7 cm in diameter, 
positioned on the beam axis  close to the center of CLAS.
The target entrance and exit windows were 15 $\mu$m  Al foils. 
A 4 cm  vertex cut for the scattered electron selected events from the central part of the target 
and eliminated events from the windows. The CLAS vertex resolution~\cite{CLAS} of $\sigma=2$ mm allowed us 
to estimate a background from the windows of $< 0.5$\%~\cite{kim2}. 

Electrons and protons from the reaction $^2$H$(e,e^\prime p)n$ were selected in fiducial regions 
of CLAS, where the particle detection efficiency is high and nearly constant.
Both the CC and EC were used to distinguish electrons from pions for
momenta $< 2.8$ GeV/c,
whereas only the EC was used for 
momenta $> 2.8$ GeV/c
where the CC became sensitive to pions.  Ref.~\cite{kim2} reports that $\pi^-$ 
contamination is 
$<2$\% 
depending on $Q^2$. The data were corrected for this effect. The protons were identified 
using tracking and time of flight~\cite{CLAS}.

The electron  detection efficiency depends on the drift-chamber
inefficiency (2.5\%) and the $\pi^-$ rejection cuts in the EC (2.5\%) and the CC (10\%), on average.
The proton   detection efficiency depends on the $\pi^+$  rejection cut (2.5\%) and the
inefficiency of  the drift chambers plus the
time-of-flight scintillators (10\%)~\cite{kim2}.  

The exclusive $^2$H$(e,e^\prime p)n$ events were extracted from the data by requiring the missing mass to be 
that of the undetected recoil neutron.
We measured the  differential $^2$H$(e,e^\prime p)n$ cross section  as a function of  $Q^2$,
$p_n$ and $\theta_n$ (the neutron polar angle with respect to the momentum transfer direction),  integrated it over $\phi_n$ (the azimuthal angle of the recoil neutron), and corrected it
for acceptance and radiative effects.
The acceptance corrections were calculated using a 
Monte Carlo technique for all $Q^2, p_n$ and $\theta_n$ bins, and were
applied event by event to every bin.   
The radiative correction factors were calculated 
using the method described in Ref.~\cite{larry}.

\begin{figure}[hbt]
\begin{center}
\epsfig{file=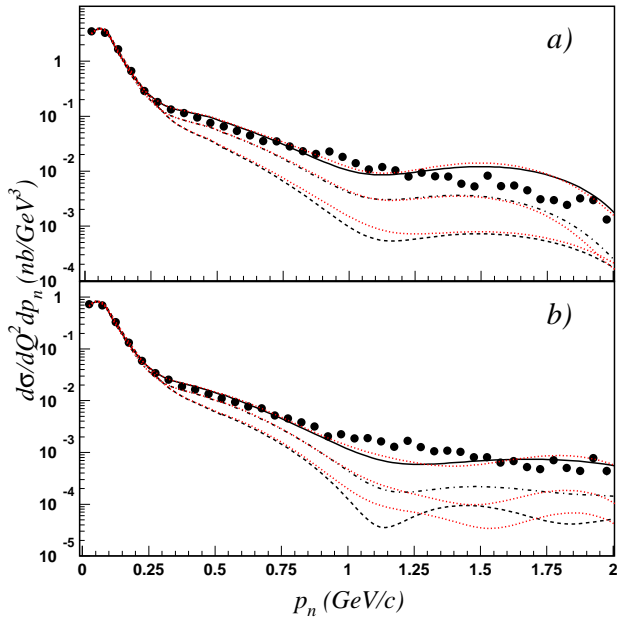, width=9.0cm, angle=0}
\caption[]{Color online. The recoil neutron momentum distribution for (a) $Q^2=2\pm0.25$~GeV$^2$ and (b) $Q^2=3\pm 0.5$~GeV$^2$.  
Dashed, dash-dotted and solid curves are calculations with the Paris potential for PWIA, PWIA+FSI 
and PWIA+FSI+MEC+N$\Delta$,  respectively.
Dotted (red) curves are calculations with the AV18 potential.}
\label{pn_23}
\end{center}
\end{figure}

\begin{figure}[hbt]
\begin{center}
\epsfig{file=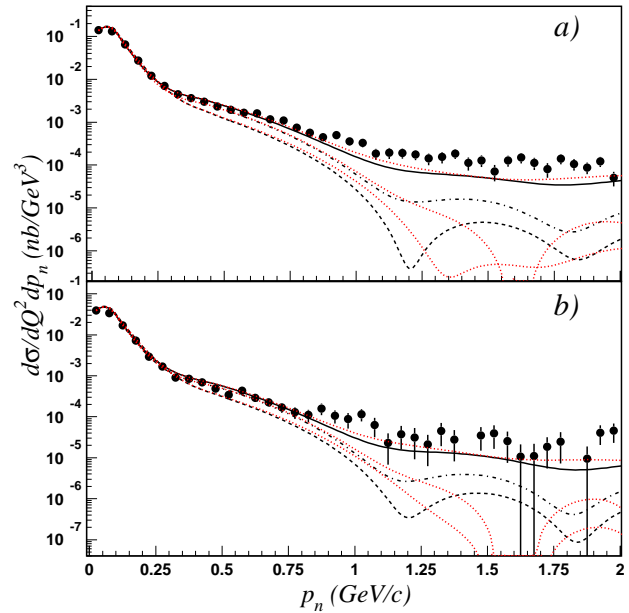, width=9.cm, angle=0}
\caption[]{Color online. The same as Fig.\ref{pn_23} for (a) $Q^2=4\pm 0.5$~GeV$^2$ and (b) $Q^2=5\pm 0.5$~GeV$^2$. }
\label{pn_45}
\end{center}
\end{figure}

\begin{figure}[hbt]
\begin{center}
\epsfig{file=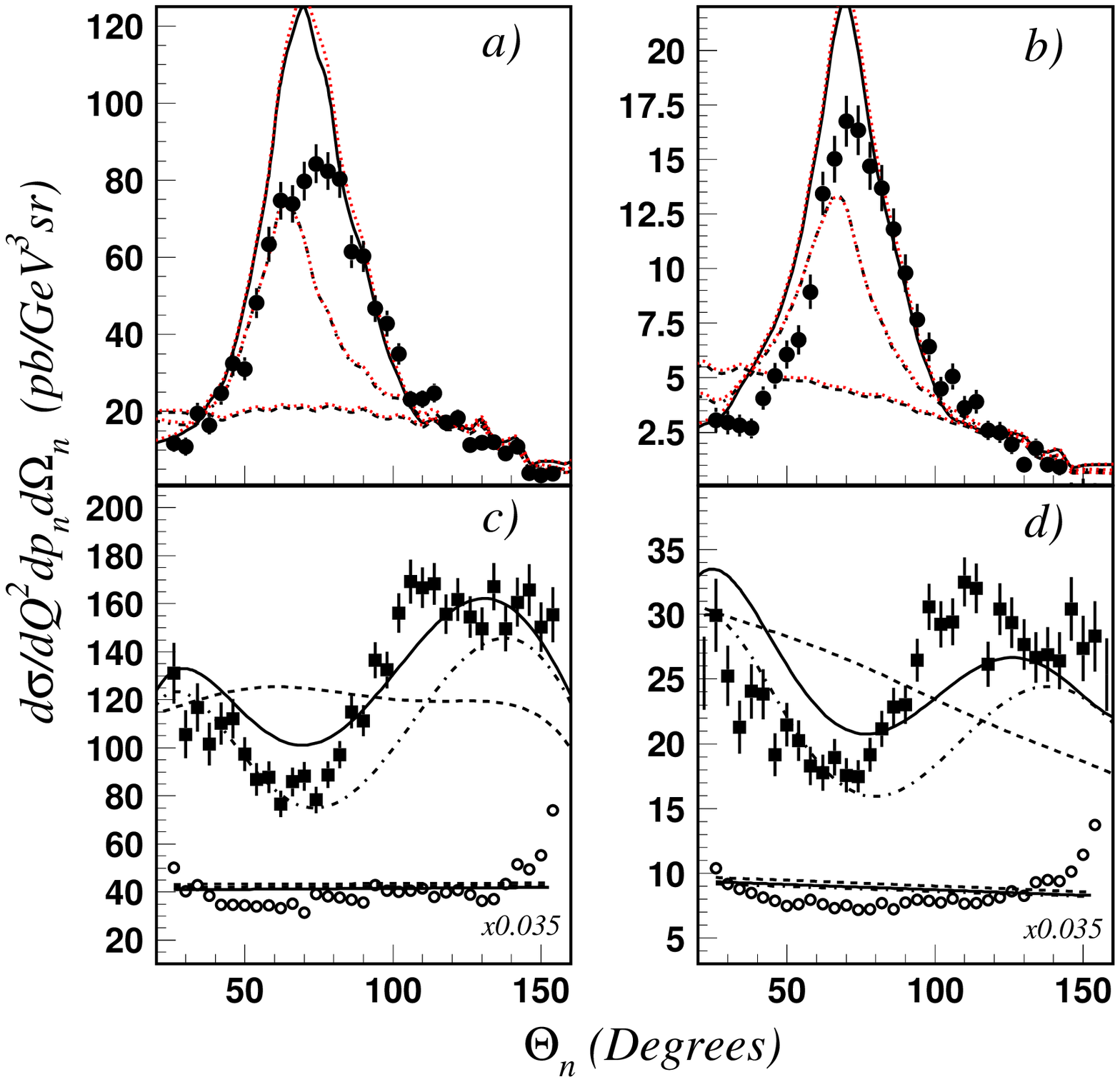, width=9.5cm, angle=0}
\caption[]{Color online. The recoil neutron angular distribution for (a) $Q^2=2\pm 0.25$~GeV$^2$, $400<p_n<600$~MeV/c; 
(b) $Q^2=3\pm 0.5$~GeV$^2$
$400<p_n<600$~MeV/c;
(c) $Q^2=2\pm 0.25  $~GeV$^2$, $200<p_n<300$~MeV/c; and (d) $Q^2=3\pm 0.5$~GeV$^2$, $200<p_n<300$~MeV/c. The data 
for $p_n<100$~MeV/c 
are plotted in the bottom part of (c) and (d) and scaled by 0.035. The curves have the same meaning as in Fig.~\ref{pn_23}.}
\label{ang_23}
\end{center}
\end{figure}

\begin{figure}[hbt]
\begin{center}
\epsfig{file=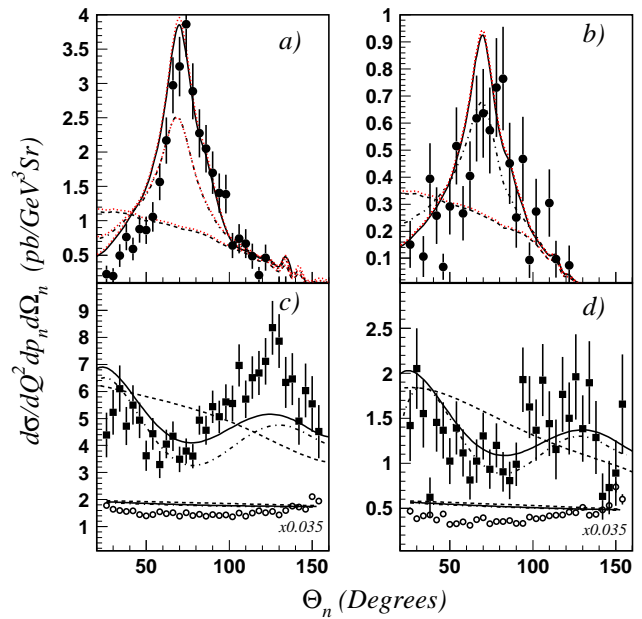, width=9.5cm, angle=0}
\caption[]{Color online. The same as in Fig~\ref{ang_23}, but for $Q^2=4\pm 0.5$~GeV$^2$ ((a) and 
(c)) and $Q^2=5\pm0.5$~GeV$^2$ ((b) and (d)).}
\label{ang_45}
\end{center}
\end{figure}

The measured  cross sections (points) are shown versus $p_n$ in Figs.~\ref{pn_23}, \ref{pn_45}, and versus $\theta_n$
in Figs.~\ref{ang_23}, \ref{ang_45} for Q$^2=$ 2, 3, 4 and 5 GeV$^2$. 
Statistical errors only are shown.
Systematic uncertainties due to the pion contamination, electron and proton detection efficiency, beam intensity measurements, and
target density  are less than 1\%.  More important are the uncertainties from the effective target length (3.5\%), 
acceptance corrections (5.5\% point to point), background subtractions from
the missing mass distributions (2-3\% average; 5.5\% point to point),  and radiative corrections  (4\%).  
 The total experimental systematic uncertainty is 10\% \cite{kim2}. 

We have investigated the same reaction theoretically using the most recent 
predictions of Ref.~\cite{La05}, which have been programmed into 
a Monte-Carlo
code that generates events in the fiducial acceptance of CLAS. We sampled $p_n$,
$\theta_n$, $\phi_n$ , $\phi_e$ (the azimuthal angle of the scattered electron) and 
$Q^2$ from a flat distribution, and then calculated all remaining momenta and angles constrained by quasi-elastic 
kinematics.  If the electron and 
the proton fell in the CLAS acceptance we recorded the kinematics of the event in a form of an Ntuple~\cite{PAW} and we weighted it with the corresponding  cross section, differential in $p_n$, 
$\theta_n$, $\phi_n$, $Q^2$ and $\phi_e$.  The events were then binned identically to the experimental data using the same cuts. 
No normalization factors between theoretical and experimental data were used. 

This model 
is an extension of earlier diagrammatic methods~\cite{La81,La94}  to JLab kinematics.
It incorporates four amplitudes: the Plane Wave Impulse Approximation (PWIA), meson exchange currents (MECs), high energy 
diffractive nucleon-nucleon elastic scattering (FSIs) and intermediate $\Delta$-nucleon rescattering ($\Delta$N).
Deuteron wave functions derived from both the Paris~\cite{PaXX} and the Argonne V18~\cite{AV18} potentials were used. 
The electron couples to the nucleons through a fully relativistic, on-shell nucleon current. 
The dipole parameterization was chosen for the magnetic form factors of the nucleon. The latest JLab data~\cite{GepXX} 
were used for the proton electric form factor,
while the Galster~\cite{GalXX} parameterization was selected for the 
neutron electric form factor. The parameters of the NN amplitude are the same as 
in Ref.~\cite{La05}, and are fixed by the elastic scattering cross section. The $\pi$ and $\rho$ exchanges are taken into account 
in the MEC  and $\Delta$N formation amplitudes, as described in Ref.~\cite{La94}. 
The electromagnetic $N\rightarrow\Delta$ transition form factor $F_{N\Delta}(Q^2)=(1-Q^2/9)/(1+Q^2/0.7)^2$ 
is driven by the world data (MAID parameterization~\cite{MAID}) and specifically by
the highest $Q^2$ measurement~\cite{DeEM} in Hall C at JLab. The most 
recent data set~\cite{Un06} from CLAS  is lower by as much as 10\%  for $Q^2< 3$~GeV$^2$ but  is similar for $Q^2>3$~GeV$^2$.

The calculated cross sections are shown in Figs.~\ref{pn_23}, \ref{pn_45}, \ref{ang_23}, and \ref{ang_45}.
Systematic uncertainties in the theoretical cross sections come from the on-shell approximation for the 
electron-nucleon current ($\sim$5\%), the parameterization of the NN elastic scattering amplitude 
($\sim$10\%), 
and  the parameterization of $F_{N\Delta}$  ($\sim$11\%).  Thus, the systematic uncertainties 
in the theoretical predictions are 
$\sim 15$\% for the full calculations.
Since the MEC amplitude in our $Q^2$ range is very small, the corresponding uncertainty can be neglected.

Figs.~\ref{pn_23} and~\ref{pn_45} show the distributions in recoil neutron momentum  integrated 
over the angular range $20^{\circ}<\theta_n< 160^{\circ}$, where acceptance 
corrections are well defined \cite{kim2}.
The experimental $p_n$ distribution drops by three orders of magnitude over the range 0--2 GeV/c
similar to the full 
theoretical calculations. For $p_n<$ 800 MeV/c, however, the data and calculations agree better
than for higher $p_n$.
Below $p_n=250$~MeV/c, quasi-elastic scattering of electrons on protons (the PWIA channel) exhausts the 
cross section.  Neutron-proton FSI dominates for $250<p_n<750$ MeV/c, 
while intermediate $\Delta$ production is prominent for $p_n>750$ MeV/c, bringing the model close to the data. Both Paris~\cite{PaXX} and Argonne V18~\cite{AV18} wave functions show similar results for $p_n<$ 1 GeV/c, 
whereas above 1 GeV/c the two wave functions differ strongly, and lead to  very different PWIA contributions. 
However, the $\Delta N$ channel overwhelms the cross section here: low momentum components of the wave function feed these higher values of $p_n$, and the sensitivity of the cross section to the high 
momentum components of the wave function is lost. Nevertheless, the theory agrees well with the data.
 
Although the theory describes the neutron momentum distributions well,
the log scale makes a close comparison difficult.
The remaining differences between theory and experiment are best seen quantitatively in the 
linear plots of angular distributions for various 
regions in $p_n$ below 600 MeV/c.
The detailed theoretical investigations in Refs. \cite{SaXX, La98} have shown that there are 
specific features of recoil neutron angular distributions for different  ranges in $p_n$.
For $p_n<  0.1$ GeV/c the angular distributions are
expected to be insensitive to FSIs,
for $p_n\sim 0.4-0.5$ GeV/c FSIs should dominate,
and for $0.2<p_n<0.3$ GeV/c the interference  between PWIA and FSI amplitudes should contribute. Until now, this characteristic behavior of the recoil neutron angular distributions has not been checked experimentally.
 
Figs.~\ref{ang_23} and~\ref{ang_45} show  neutron angular distributions for three ranges of $p_n$ at $Q^2=2, 3, 4$ and $5$ GeV$^2$. Each panel clearly shows the evolution of the interaction effects with $p_n$ and $\theta_n$, for a fixed value of $Q^2$. 

In the highest momentum range ($0.4<p_n<0.6$ GeV/c) the angular distributions exhibit a large peak in the vicinity 
of $\theta_n= 70^{\circ}$. 
This effect comes from neutron-proton rescattering, and corresponds to the on-shell propagation 
of the struck  nucleon. It is 
maximal when the kinematics allow for rescattering on a nucleon almost at rest \cite{La81},
which happens when $x=Q^2/2M\nu=1$ ($\nu$ is the energy of the virtual photon, and $M$ is the 
nucleon mass). The following physical picture emerges. The electron scatters primarily from a proton almost at rest. Since the total
energy is larger than the sum of the masses of the two nucleons, the struck proton can propagate on-shell and rescatter off the 
neutron which is also nearly at rest. In the lab frame, the soft neutron recoils at 90$^{\circ}$ with respect to the fast 
forward proton. Two-body kinematics places the rescattering peak at about $\theta_n=$70$^{\circ}$ for our kinematics.
In the classical Glauber approximation, the nucleon propagator is linearized and recoil effects 
are neglected, and herefore, the rescattering peak stays at $\theta_n=$90$^{\circ}$~\cite{Je96,Cio01}. This has been fixed in the 
Generalized Eikonal Approximation (GEA)~\cite{SaXX} which takes into account higher order recoil terms in the nucleon propagator. In the diagrammatic approach 
the full kinematics are taken into account from the beginning~\cite{La81,La05}. The shape of the angular distribution reflects 
the momentum 
distribution of the proton in deuterium.

A $\Delta$ resonance  produced on a nucleon at rest at $x=[1+(M^2_{\Delta}-M^2)/Q^2]^{-1}<1$, can propagate 
on-shell and rescatters from the second nucleon also at rest~\cite{La05}. 
This contribution shifts the rescattering peak toward larger angles, and 
brings the theory  into better agreement with experiment. It also decreases faster with $Q^2$, consistent with the steeper  
variation  of the 
$N\rightarrow \Delta$ transition electromagnetic form factor as compared to the dipole parameterization of the nucleon form 
factors. The excess theoretical cross section at $Q^2=2$~GeV$^2$ is a reflection of our linear fit to the ratio of  $N\rightarrow\Delta$ 
and dipole form factors. A better fit to the latest data~\cite{Un06} from CLAS leads to a reduction of the peak by $\sim 15$~\% 
for $Q^2<3$ GeV$^2$, in better agreement with experiment.

In the intermediate momentum range ($0.2<p_n<0.3$ GeV/c), FSIs suppress the quasi-elastic contribution in the vicinity of $x=1$. 
Here the relative kinetic energy  between the outgoing proton and neutron $T\sim Q^2/2M$ lies between 1 and 3~GeV. The nucleon-nucleon scattering amplitude is almost purely absorptive and the FSI amplitude interferes destructively with the quasi-free amplitude.  This induces a loss of flux for fast protons.

In the lowest momentum range  ($p_n<0.1$ GeV/c)  rescattering effects are small, and the experimental and theoretical 
angular distributions are similarly flat. 
The magnitude of the experimental cross sections is well reproduced at low $Q^2$, but  
the theory slightly exceeds the data at larger $Q^2$. This effect has already been observed in the study of 
$^3$He$(e,e^\prime p)^2$H 
at low recoil momentum~\cite{VoXX}, and is not yet well understood.

In summary, this benchmark experiment demonstrates that the mechanisms of the exclusive $^2$H$(e,e^\prime p)n$ reaction  
 are well understood for $1.75<Q^2<5.5$ GeV$^2$. Theoretical and experimental cross sections agree within  20\%,  
consistent with the 
systematic uncertainties ($\approx$15\% for theory and $\approx$10\% for experiment).  
Proton-neutron rescattering (FSIs) and $\Delta$ production dominate over a large part of the phase space, except at 
backward angles 
($\theta_n > 110^{\circ}$) or  very low recoil momenta ($p_n < 100$ MeV/c), where the distributions directly reflect 
the deuteron wave function.
A good understanding of the mechanisms of the exclusive $^2$H$(e,e^\prime p)n$ reaction in our $Q^2$ region opens an opportunity
to investigate the short distance properties of nucleons in deuterium, which will be discussed in our future publications.

We thank the staff of the Accelerator and Physics Divisions at Jefferson Lab for their support. 
This work was supported in part by the U.S. Department of Energy (DOE), the National Science Foundation, 
the Armenian Ministry of Education and Science, the French Commissariat \`a l'Energie Atomique, the French Centre National de la 
Recherche Scientifique, the Italian Istituto Nazionale di Fisica Nucleare, and the Korea Research Foundation.
Authored by The Southeastern Universities Research Association, Inc. under U.S. DOE Contract No. DE-AC05-84150. The U.S. Government retains a non-exclusive, paid-up, irrevocable, world-wide license to publish or reproduce this manuscript for U.S. Government purposes.

\end{document}